# Optical frequency analysis on dark state of a single trapped ion

**ADAM LEŠUNDÁK[1],[\*], TUAN M. PHAM[1], MARTIN ČÍŽEK[1], PETR OBŠIL[2], LUKÁŠ SLODIČKA[2], AND ONDŘEJ ČÍP[1]**

[1] *Institute of Scientific Instruments of the Czech Academy of Sciences, Královopolská 147, 612 64 Brno, Czech Republic*
[2]*Department of Optics, Palacký University, 17. listopadu 12, 771 46 Olomouc, Czech Republic*
*\*lesundak@isibrno.cz*

**Abstract:** We demonstrate an optical frequency analysis method using the Fourier transform of detection times of fluorescence photons emitted from a single trapped $^{40}Ca^+$ ion. The response of the detected photon rate to the relative laser frequency deviations is recorded within the slope of a dark resonance formed in the lambda-type energy level scheme corresponding to two optical dipole transitions. This approach enhances the sensitivity to the small frequency deviations and does so with reciprocal dependence on the fluorescence rate. The employed lasers are phase locked to an optical frequency comb, which allows for precise calibration of optical frequency analysis by deterministic modulation of the analyzed laser beam with respect to the reference beam. The attainable high signal-to-noise ratios of up to a MHz range of modulation deviations and up to a hundred kHz modulation frequencies promise the applicability of the presented results in a broad range of optical spectroscopic applications.

## 1. Introduction

The optical atomic spectroscopy is a well-developed research field comprising some of the most advanced applications of optical and laser technologies. It provides many crucial investigation methods and contributes to a whole spectrum of modern natural sciences. At the same time, it still holds an immense potential for further advancements by utilization of newly available tools for light generation, control and analysis developed within studies of light and matter at the level of individual particles [1-3]. The possibilities of employing the individual atomic systems for optical spectroscopy have in the last decade to a large extent governed the development in the field of optical frequency metrology and its attainable accuracies [4]. The implementation of frequency analysis with individual atomic particles has been dominantly focused on the development of optical frequency references utilizing a direct probing of narrow atomic transitions. Related phenomenological advancements have mostly been corresponding to knowledge improvements of atomic internal structure and its sensitivity to external disturbances. The development of advanced methods for laser frequency stabilization complemented by progress in generation and control of optical frequency combs seen in the past few years have recently enabled the pioneering excitations of Raman transitions in the gigahertz [5, 6] and terahertz domain [7] as well as new atomic spectroscopy methods [8].

Here we present a method for optical frequency analysis based on the time resolved spectroscopy on atomic dipole transitions with enhanced sensitivity to frequency deviations using the probing of individual atoms close to a two-photon resonance in a lambda-like three-level energy scheme. The spectral resolution and bandwidth of the presented approach profit from both, the high attainable count rates on atomic transitions with large decay rates and, at the same time, possibilities of continuous probing of narrow two-photon resonances with disparate optical wavelengths enabled by the sub-wavelength localization of probed single atom. The two-photon interference not only enhances the spectral sensitivity but, crucially, it allows for optical frequency analysis of target optical field on optical reference field with very different frequency, which corresponds to a paramount task in many metrological applications [9-11]. The interference mediated by laser cooled atoms allows for optical frequency beating of two optical fields separated by hundreds of THz. This is typically realized indirectly via an optical frequency comb, which provides a traceability to stable frequency reference in a broad optical ruler [12, 13]. The presented method provides an alternative approach within limited spectral range, by representing a convertor of the relative optical frequency difference to the intensity of fluorescence emitted by a single atom.

We implemented a continuous measurement scheme on a single trapped $^{40}Ca^+$ ion excited in the lambda-type energy level scheme $4S_{1/2} \leftrightarrow 4P_{1/2} \leftrightarrow 3D_{3/2}$ by two lasers at wavelengths 397 nm and 866 nm, respectively. The 397 nm laser is simultaneously used for Doppler cooling, while the 866 nm laser serves as a repumper from a long-lived $3D_{3/2}$ manifold. Both lasers are phase locked to particular teeth of a fiber frequency comb referenced to a hydrogen maser. In this way, they adopt known frequency linewidths and stability of the optical reference. We analyze the properties of the presented scheme in terms of the fluorescence intensity response to imposed laser frequency characteristics and measurement length. For this purpose, the repumping laser serves as an analyzed field to which a deterministic frequency modulation is applied, while the cooling laser is the reference field with fixed frequency. A simple theoretical model for achievable signal-to-noise ratios (*SNR*) is compared with measured fluorescence intensity responses in a range of modulation frequencies and amplitudes. Finally, we estimate the limits of the presented method and discuss the comparisons of measured data with simulations.

## 2. Experimental scheme

To measure the dynamic response of ion fluorescence to the frequency detuning of the excitation lasers, it is necessary to narrow the linewidths and stabilize the frequencies of the lasers to levels where they do not interfere with the measurements. For this reason, two extended-cavity diode lasers (ECDLs) are phase locked to the optical frequency comb. The whole experimental setup is schematically shown in Fig. 1 with three parts: a) is dedicated to the ECDLs, b) to the phase locking and c) to the ion trap. The frequency comb with a center frequency at 1550 nm is frequency doubled by the second harmonic generation (SHG) process and broadened by a photonic crystal fiber to generate an optical supercontinuum ranging from about 600 nm to 900 nm. This broad spectrum is mixed with both laser beams to create beat notes with particular frequency comb teeth. The product of this mixing is spatially separated according to the wavelengths by a diffraction grating and serves as a signal for phase locked loop (PLL), which efficiently narrows a laser linewidth down to linewidth of a single component of the frequency comb. This method allows for stabilization of multiple lasers in near infrared and visible range to a single frequency reference within one optical setup. The two radio frequency parameters of optical frequency comb - the repetition rate $f_{rep}$ and the offset frequency $f_{ceo}$ are referenced to a hydrogen maser, which is disciplined for long-term stability by a GPS clock.

Particular beat note signal between the nearest comb tooth and the laser is amplified, and mixed with a reference signal. This reference signal is generated by a radiofrequency (RF) synthesizer referenced to the hydrogen maser, and has the value of desired beat note frequency $f_b$. After low-pass filtering, the product of the mixing is a DC signal that is proportional to the phase error of the comb-laser beat note and serves as the error signal for laser phase-locked loop (PLL). The PLL is realized with fast analog control electronics. In this way, the laser linewidths efficiently adopt the linewidths of the frequency comb. The cooling laser at 397 nm cannot be directly locked to the frequency comb and the lock is realized at its fundamental wavelength at 794 nm. The laser frequencies written in terms of the comb frequencies are then: $v_{866} = 2(n_{(2 \cdot 866)}f_{rep} + f_{ceo}) + f_{b866}$ and $v_{397} = 2[2(n_{(2 \cdot 794)}f_{rep} + f_{ceo}) + f_{b794}]$. The frequency difference between lasers as seen by the ion is $\Delta_\nu = (4n_{(2\cdot794)} - 2n_{(2\cdot866)}) f_{rep} + 2f_{ceo} + f_{b794} - f_{b866}$, where $4n_{(2\cdot794)} - 2n_{(2\cdot866)} = 1635856$. Standard deviations of the comb basic frequencies are $\sigma(f_{rep}) \leq 1$ mHz and $\sigma(f_{ceo}) \leq 1$ Hz at 100 s averaging time, which indicates standard deviation of laser frequency detuning below 1 Hz at the same averaging timescales. We measured the linewidth of a single comb tooth by beating it with a laser at 1540 nm which has linewidth at the level of a few Hz and obtained FWHM = 40 kHz with 1 kHz resolution bandwidth setting. The main contribution to this linewidth comes from the $f_{ceo}$. The $f_{rep}$ contribution in these time scales can be neglected. After each SHG process, the linewidth is also doubled. However, since the noise source is common for both 397 nm and 866 nm lasers, a simple estimation of their mutual linewidth based on $\Delta_\nu$ leads to 80 kHz.

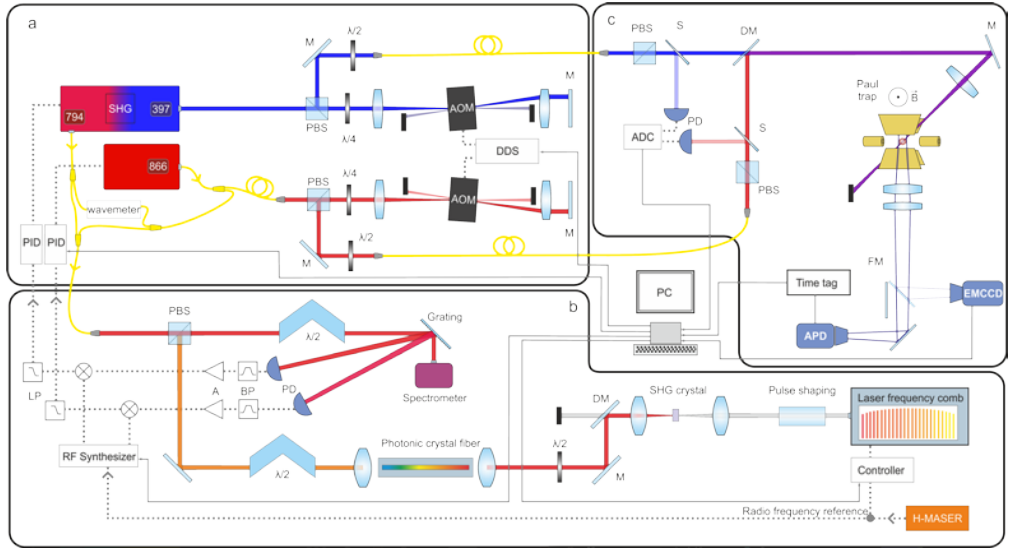

Fig. 1. Scheme of the experimental setup distributed over three optical tables corresponding to parts *a, b* and *c*. The cooling laser beam is generated by doubling the 794 nm ECDL laser to 397 nm. The near infrared laser beams are sent through an optical fiber to table *b* for frequency stabilization. Laser beams prepared for excitation with the trapped ion are led into acousto-optic modulators (AOMs) in double pass configurations for frequency and amplitude modulation and then sent to the ion trap positioned on optical table *c*. The optical frequency comb comprising the units for frequency doubling and supercontinuum generation is optically mixed with lasers for generating the error signals for frequency stabilization. The excitation laser fields are set to particular desired polarization and are amplitude stabilized at the proximity of the ion chamber on optical table *c*. The single ion fluorescence is collected by a lens with numerical aperture of 0.2 and detected by EMCCD camera or single photon avalanche diode (APD). The detection signals including precise photon arrival times are recorded using fast time-tagging module or converted by analog-digital card (ADC) for further processing.

The target frequencies of atomic transitions are achieved by fine frequency tuning with acousto-optic modulators (AOM) in double-pass configuration. These AOMs are also used for stabilization of excitation laser intensities. The beams are combined by a dichroic mirror and focused to the center of the linear Paul trap with 45° angle of incidence relative to the trap axis. The ion is trapped in the radial *x, y* directions by harmonically oscillating electric field at frequency of 30 MHz and with an amplitude corresponding to the radial secular motion frequency $f_{x,y} \approx 1.66$ MHz. The axial position is confined by the voltage applied to tip electrodes $U_{tip} = 500$ V, leading to the axial secular frequency $f_z \approx 780$ kHz. The ion is Doppler cooled by a red detuned 397 nm laser. Fluorescence from the ion is collected in the direction of the magnetic field using a lens system with numerical aperture of 0.2. A flip mirror in the optical path is used to switch the detection between the electron multiplying CCD (EMCCD) camera and single-photon avalanche detector (SPAD). The photon detection times are recorded with fast time tagging module with up to 4 ps resolution.

In the case of $^{40}$Ca$^+$ ion, the convenient three-level systems with Λ configuration can be realized on transitions $4S_{1/2} \leftrightarrow 4P_{1/2} \leftrightarrow 3D_{3/2}$, which are simultaneously driven by two laser fields at wavelengths 397 nm and 866 nm. For the individual laser detunings set to the two-photon resonance corresponding to the vanishing difference $\Delta_{866} - \Delta_{397} = 0$, the population of the excited state $4P_{1/2}$ disappears in ideal case. This corresponds to a dark state with no fluorescence emission [14, 15]. In practice, the dark state population is limited by the finite linewidth of the two involved lasers and finite coherence between the ground states $4S_{1/2}$ and $3D_{3/2}$ as well as thermal population of atomic motion [16-19]. The magnetic field of 6.1 Gauss is applied at the position of the ion to lift the degeneracy of Zeeman states which results in effective internal energy level scheme with eight states, see Fig. 2-a). The angle between linear polarization of laser fields is chosen to be perpendicular to the magnetic field, therefore only transitions with $\Delta m_j = \pm 1$ are excited, which allows for efficient depopulation of the outermost states from the $3D_{3/2}$ manifold and leads to the observation of four dark resonances. Tuning the 866 nm laser to the slope of particular dark resonance, which is described by the slope parameter $m(\Delta_{866})$ representing the resonance gradient, allows to use the ion as a direct convertor of frequency deviation to fluorescence intensity due to its quasi-linear dependence. See Fig. 2-b) for schematic explanation. The fluorescence spectrum with dark resonances can be very well reproduced using standard approach based on 8-level optical Bloch equations [20], which allow for the estimation or cross-check of several experimental parameters including laser detunings, intensities or magnetic field amplitude and direction.

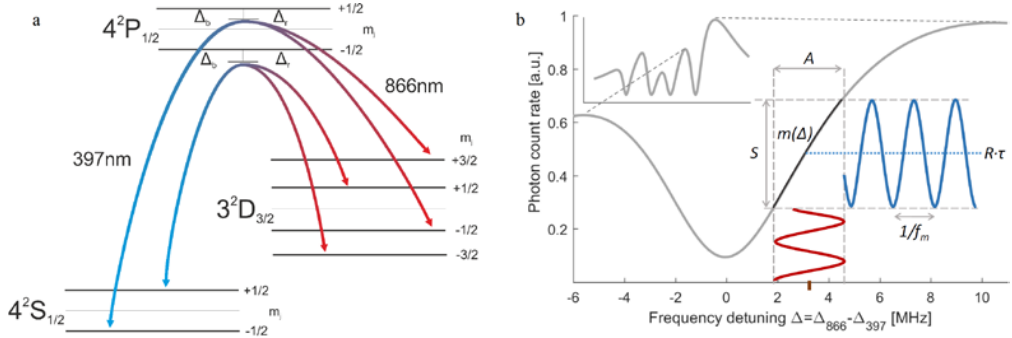

Fig. 2. Principle of operation of time-resolved optical spectroscopy close to a dark-resonance of a trapped $^{40}$Ca$^+$ ion. a) The employed energy level scheme of $^{40}$Ca$^+$. The two laser fields with polarizations perpendicular to the magnetic field lead to observation of four dark states between $S_{1/2}$ and $D_{3/2}$ manifolds. b) A simulation of the $P_{1/2}$ level occupation probability proportional to fluorescence intensity as a function of 866 nm laser frequency detuning $\Delta_{866}$. The detail of the selected dark resonance depicts the fundamental parameters determining the performance of the spectral analysis including the detuning $\Delta_{866}$, frequency deviation of the frequency modulation $A$, modulation frequency $f_m$, conversion through the slope with parameter $m$ onto the fluorescence with an average photon rate per second $R$ and measured with a SPAD gate interval $\tau$.

The enhancement of the dark resonance contrast and slope steepness by utilization of frequency stabilization of 397 nm and 866 nm lasers to the frequency comb is illustrated in Fig. 3. At this locking configuration, the fluorescence spectrum is measured, though only in narrow bandwidth limited by the PLL frequency range corresponding to $f_{rep}/2$. The whole spectrum containing all dark resonances is measured again but with the 866 nm laser frequency locked to the wavemeter with time constant in order of a second. This "wavemeter frequency lock" does not enhance the coherence properties of the laser in relevant frequency bandwidth in our scope of interest, but ensures well defined frequency scanning of the whole dark resonance spectra. Both fluorescence spectra are fitted with the optical Bloch equations; see Fig. 3. From observed four resonances, one corresponding to transition between electronic states $|S_{1/2},+1/2\rangle$ and $|D_{3/2},-3/2\rangle$ is chosen for frequency analysis experiments because it is the closest one to a global fluorescence maximum and has the longest resonance slope. The measurement point $\Delta_{M1}$ is chosen in the middle of the slope of the dark resonance to maximize the measurable amplitudes of frequency modulation. The slope parameter for linear approximation at this measurement point was evaluated to $m(\Delta_{M1}) = 1.79\pm0.01$ counts s$^{-1}$·kHz$^{-1}$, with corresponding photon count rate of 6800 counts·s$^{-1}$. The frequency detunings of analyzed and reference lasers from their corresponding transition frequencies are $\Delta_{866} \approx -9$ MHz and $\Delta_{397} \approx -12$ MHz. The fit of optical Bloch equations of the measured fluorescence rates does not perfectly follow the fluorescence curve since the equations do not include the ion motion affected by the Raman cooling and heating processes naturally involved within the measurement scheme [16]. Particularly at the measurement point $\Delta_{M1}$ its gradient corresponds to $m_{Bloch}(\Delta_{M1}) = 2.21$ counts·s$^{-1}$·kHz$^{-1}$, which is 24% greater than the gradient of the numerical fit. For this reason, we use the polynomial fit as the slope function $m_{nl}$, instead of the fit of the optical Bloch equations. The fit parameters are: the magnetic field $\mathbf{B} = 6.1$ Gauss, the detuning of the blue laser $\Delta_{397} = -12$ MHz, angle between the light fields and the magnetic field $\alpha= 90°$, saturation parameters $S_{397}=1$, $S_{866}=4$ and combined linewidth of the two lasers $\Gamma = 251$ kHz and $\Gamma=124$ kHz for the case of employed PLL frequency stabilization.

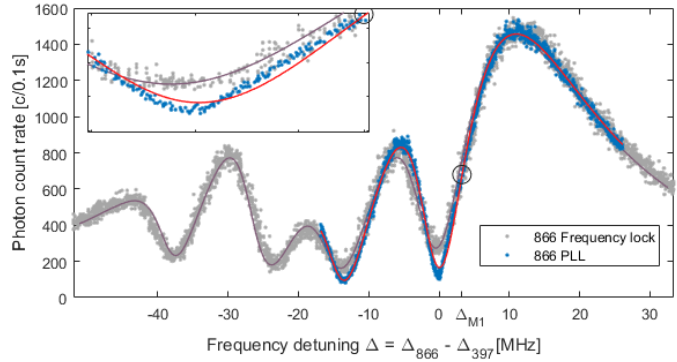

Fig. 3. Dark blue dots represent the fluorescence spectrum measured with both ECDL lasers phase-locked to the frequency comb with limited bandwidth due to PLL limited scanning range. Red line is the fit and black circle corresponds to the measurement point. Parameters of the fit are the magnetic field: $\mathbf{B}= 6.1$ Gauss, the detuning of the blue laser $\Delta_{397} = -12$ MHz, angle between the light fields and the magnetic field $\alpha= 90°$, saturation parameters $S_{397}=1$, $S_{866}=4$ and combined linewidth of both lasers $\Gamma=124$ kHz. Light blue dots represent the fluorescence spectrum with all dark resonances visible, measured by scanning of the 866 nm laser frequency locked to the wavemeter. Purple line shows its theoretical fit with the lasers combined linewidth parameter $\Gamma=251$ kHz. The black circle marks the chosen measurement point $\Delta_{M1}$. The inset depicts the lower part of the resonance slope to show the differences between the two spectra and the fits in the relevant region.

## 3. Analysis of the spectral sensitivity

To determine the dynamic response of the ion as a detector of mutual frequency shifts of the two lasers, we introduce a controlled frequency modulation of one of the lasers followed by fluorescence signal processing. A phase noise power spectral density distribution of both phase locked ECDLs to the optical frequency comb has dominantly $1/f$ noise shape profile [21]. The low magnitude of the noise has no influence on the dynamic response of the ion characterization. The modulation is applied to the 866 nm repumping laser as a harmonic frequency modulation with specific modulation frequency $f_m$ and peak frequency deviation $A$ (frequency deviation), see Fig. 2. The modulation results in a periodic signal on fluorescence which can be considered harmonic due to the quasi-linearity of the slope close to the measurement point. The fluorescence signal is detected by the SPAD and recorded as arrival times of detected photons by the time tagging module. The timing resolution of the detected fluorescence is limited by the resolution of the SPAD and is on the level of 1 ns. The processing of such discrete signal with a nanosecond-sampling period can be simplified by summing the number of photon counts over time $\tau$ (gate interval). This acts as the filtering by a low-pass filter, which does not limit the extracted information about the spectrum if the fluorescence signal does not contain significant spectral components above the low-pass filter frequency band. The time tag record is thus processed into data of photon count rate in units of number of photon counts per interval bin $\tau$.

To determine the theoretical response function of the detected fluorescence photon rate to the introduced frequency modulation, we compare natural fluorescence noise to the observable signal in the evaluated frequency spectrum. The equation for the SNR is derived as the ratio of function describing the amplitude of harmonically modulated fluorescence $S(f_m,\tau,A)$ to amplitude of fluorescence detection noise $N(R,T,\tau)$. The signal function converts the 866 nm laser frequency deviation $A$ to the amplitude of fluorescence harmonic signal $S(f_m,\tau,A)$ according to the slope parameter of dark resonance $m(\Delta_{866})$, modulation frequency $f_m$ and the gate interval $\tau$. Although not necessary, a linear approximation of the slope parameter can simplify the evaluation and is defined as $m=\partial(\langle n \rangle \cdot s^{-1})/\partial \Delta_{866}$ at measurement detuning point $\Delta_M$, where $\langle n \rangle \cdot s^{-1}$ is an average number of detected photons per second. In case of $(1/f_m) \ll \tau$ and the linear slope parameter $m$, the fluorescence signal amplitude can be approximated as $S(f_m,\tau,A) \approx mA\tau$. The modulation frequency $f_m$ and the gate interval $\tau$ affects effective fluorescence modulation amplitude $A_{eff}$ by averaging the fluorescence over time $\tau$,

$$A_{eff}(\tau) = \frac{1}{\tau}\left|\int_0^\tau A\cos(f_m t)\,\mathrm{d}t\right| = A\left|\mathrm{sinc}(f_m \tau)\right|. \tag{1}$$

The resulting amplitude $S$ of the imposed signal is

$$S = mA\tau\left|\mathrm{sinc}(f_m \tau)\right|. \tag{2}$$

The SNR can be then found by comparing the amplitudes of detectable signal and the noise for a given frequency bandwidth. In an ideal case, when the noise contributions from the reference laser field and the applied magnetic fields are negligible, the noise of detected fluorescence corresponds to the shot noise. The noise amplitude can then be expressed as the shot noise of the average photon count rate $R$ over number of measurement samples $n_s=T/\tau$,

$$N = \sqrt{\frac{R\tau}{T}}, \tag{3}$$

where $R=\langle n \rangle/\tau$ is the average count rate per gate interval $\tau$, and $T$ is the measurement time. The corresponding SNR is

$$SNR = \frac{S}{N} = \sqrt{\frac{T\tau}{R}}mA\left|\mathrm{sinc}(f_m \tau)\right|. \tag{4}$$

In case of large frequency deviations ($A \approx$ hundreds of kHz) it is important to include a function, which reflects a true response of the observable fluorescence rate to the relative frequency detuning of analyzed laser $\Delta_{866}$. The nonlinear slope function $m_{nl}(\Delta_{866})$ is taken from a polynomial fit of a measured fluorescence spectrum around the measurement point. To obtain the signal amplitude, modulation amplitude factor $m \cdot A$ is replaced with one half of the fluorescence difference at the two extremes of modulated detuning frequency $\Delta_M + A$ and $\Delta_M - A$. This yields

$$SNR = \frac{1}{2}\sqrt{\frac{T\tau}{R}}\sum_{n=1}^{5}\left[p_n(\Delta_M + A)^{n-1} - p_n(\Delta_M - A)^{n-1}\right]\left|\mathrm{sinc}(f_m \tau)\right|, \tag{5}$$

where $p_n$ are coefficients of $n^{\text{th}}$ degree polynomial fit $m_{nl}(\Delta_{866})$.

## 4. Measurements of the fluorescence intensity response to the laser frequency modulation

The spectrum of the analyzed 866 nm laser deviations is measured by keeping the reference 397 nm laser at a constant detuning $\Delta_{397}$ and modulating the frequency detuning $\Delta_{866}(t) = \Delta_M + A\cos(f_m t)$. The measurements are done for various sets of modulation frequencies, frequency deviations and frequency detunings. Each measurement is analyzed in terms of its frequency spectrum with respect to the time bin length $\tau$. For the given $\tau$, the FFT algorithm evaluates the spectrum and the frequency component at the given modulation frequency (or with the highest amplitude) is compared to the average amplitude of the whole spectrum. The length of individual measurements is set to be an integer multiple of $1/f_m$, to avoid the spectral leaking in the FFT spectrum. The natural unmodulated fluorescence has Poissonian distribution, thus its spectrum has the character of white noise and can be averaged as a whole.

Relatively low photon count rate does not theoretically place any fundamental limit on the detectable modulation frequency. The information about frequency modulation is obtained from the modulation of the photon counts in the time bins and the information is thus still preserved even when the average count per bin is below one. Note that this should not be confused with the frequency detuning out of dark resonance slope, e.g. detuning into the bottom of the dark resonance where the count rate could eventually drop to zero. As mentioned above, the slope function $m_{nl}$ is defined in a certain region of the count rate (2000 c·s$^{-1}$ to 14000 c·s$^{-1}$). These boundaries obviously scale with the gate interval. A check that fluorescence has not crossed these boundaries, e.g. due to laser-locking dropout or ion vacuum impurity kick, has been performed for all measurements with 10 ms gate interval. Very rare but possible dropouts with prompt reappearing on shorter time scales can be considered to have negligible effect on the resulting signal.

We summarize the results of the frequency response measurements in four figures. Three of them are done at measurement point $\Delta_{M1}$ and emphasize the $SNR(\tau)$ dependence on the critical measurement and evaluation parameters with respect to the gate interval $\tau$, simulating variable gate time of a photodetector. These important parameters include modulation frequency, frequency deviation, and measurement time. The minimum of $\tau$ is chosen such that $\text{sinc}(f_m \tau_{min}) > 0.99$ and the maximum is chosen arbitrary but always much higher than $1/f_m$. The last figure compares the $SNR(\Delta_M)$ for two modulation frequencies with respect to the measurement point $\Delta_M$ to allow for estimation of optimal measurement points in future experiments. Note that we do not attempt to cover the whole accessible spectral range of modulation frequencies $f_m$ and frequency deviations $A$, but rather illustrate the working mechanism of the presented scheme and show its intrinsic limits.

### 4.1 Modulation frequency

First, we evaluate $SNR(f_m, \tau)$ performance of the method in term of the frequency bandwidth by realization of the measurements with varying modulation frequencies and constant frequency deviation. We set the length of the measurements to $T = 300$ s, frequency deviation to $A = 300$ kHz to be well within the employed resonance slope and range of modulation frequencies from $f_m = 66$ Hz to 120 kHz. For values of $\tau$ longer than the period corresponding to the modulation frequency $f_m$, the modulation component is aliased in the spectrum at aliased frequencies $f_{alias} = |b/(2\tau) - f_m|$, where folding factor $b$ is the closest even integer of multiple $2f_m\tau$. The values of signal-to-noise ratio are then $SNR = S_m/N_m$, where $S_m$ is the FFT frequency component at the modulation frequency $f_m$ or at the frequency $f_{alias}$ (for $\tau > 1/f_m$) and $N_m$ is the average amplitude of the whole FFT spectrum excepting the DC component. Importantly, the single ion based spectral analysis presented in Fig. 4. shows high attainable $SNR$ in the whole measured spectral range of $f_m$. The observed $SNR(f_m, \tau)$ ratios are well reproduced by the theoretical predictions of eq. (5) up to modulation frequencies comparable to the photon count rate $R = 6800$ c·s$^{-1}$.

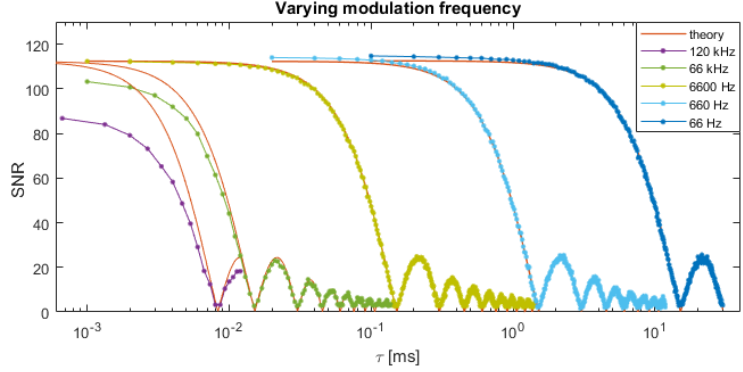

Fig. 4. Measured signal-to-noise ratio for 300 s long measurements of fluorescence intensity response to modulated detuning $\Delta_{866}$ as a function of the analysis gate interval $\tau$. Frequency deviation is kept at 300 kHz and modulation frequency varies in a broad range from 66 Hz to 120 kHz. The measured data are shown as full circles and red lines are corresponding theoretical plots evaluated using nonlinear slope function $m_{nl}$. The mean time between two successive photon detections is ~ 0.15 ms.

*4.2 Frequency deviation and measurement time*

We have studied the potential of the presented spectrometry method with respect to capturing the frequency deviation $A$ within the measurement time $T$. The expectable upper limit will be given by the spectral width of the dark resonance slope, while the lowest detectable modulation depths will depend on the amount of the detection noise. The *SNR* dependence on the gate interval is measured for low modulation frequency $f_m = 66$ Hz to avoid any effect of fast modulation on the observability of high modulation amplitudes. Figure 5. shows measured data of *SNR* for $A$ ranging from 10 to 1000 kHz and $T$ from 2.5 to 500 s.

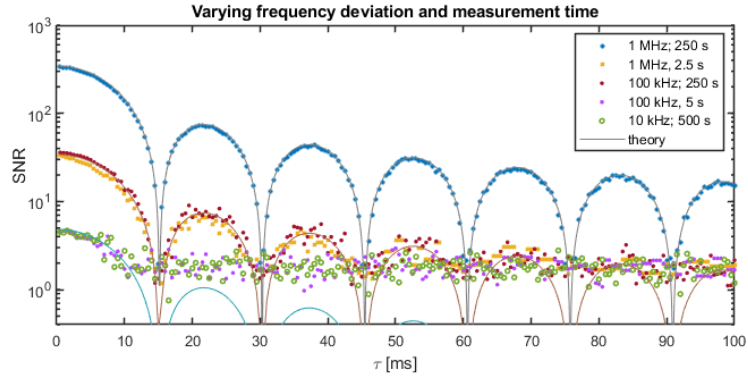

Fig. 5. Signal-to-noise ratio measurements of fluorescence with harmonically modulated detuning of analyzed laser field $\Delta_{866}$ with modulation frequency set to 66 Hz as a function of a gate interval $\tau$. The measured data for selected combinations of frequency deviations and measurement times are shown with dots. Lines show the theoretical simulation of *SNR*. Amplitude levels for signals are taken from known frequency position in the spectrum.

The combinations are chosen in a manner to advert that an $x$-times lower $A$ gives the same value of *SNR* if the measurement time is an $x^2$-times longer, reflecting the close to ideal Poissonian character of the detection noise. The measured *SNR* data nicely reproduces the theoretical predictions down to a plateau of the lowest detectable signal, which has its limits given by amplitudes of shot noise at the specific frequency components $f_m$ or $f_{alias}$. The plateau is measured to be on the level of $SNR_{pl} = 1.8 \pm 0.4$, which corresponds to $A_{pl} = 38 \pm 8$ kHz for $T = 5$ s respectively $A_{pl} = 3.8 \pm 0.8$ kHz for $T = 500$ s.

## 4.3 SNR limit for unknown modulation frequency

For a general task of estimation of amplitude of unknown frequency components, the whole FFT spectrum has to be searched. In this case, the signal-to-noise ratio can be calculated as $SNR(T,\tau) = S_{max}/N_m$, where $S_{max}$ is the highest amplitude component of the whole FFT frequency spectrum excepting the DC component. Thus, there is always a spectral component with amplitude at least on the level corresponding to the detectable limit defined here as variable $SNR_{lim}$. That means, if there is an unknown frequency modulation, we are able to detect it only if its $SNR$ is higher than $SNR_{lim}$. It is obtained by simulating data of shot noise with Poissonian distribution. The simulated data represent pure fluorescence without any modulation and have the same count rate $R$ and measurement length $T$ as the real data. This simulated fluorescence is analyzed in the same way as the measured data and the $SNR_{sim}$ is calculated. An average of 500 simulations $SNR_{sim}$ is used for estimation of the detection limit $SNR_{lim}(T,\tau) = \langle SNR_{sim} \rangle_{500}$. Fig. 6. shows two measurements of $SNR(T,\tau)$ both with $f_m = 66$ Hz, $A = 300$ kHz and with total measurement times $T$ equal to 3 and 300 s. Observed $SNR(T,\tau)$ values strictly follow theoretical curves in all regions above the limit of $SNR_{lim}(T,\tau)$. Importantly, the figure also shows the effect of the measurement length $T$ on the detectable limit $SNR_{lim}$ and as can be seen, the limit is not constant with $T$. A simple explanation is, as there is a higher probability for higher noise amplitude in longer measurements, $SNR_{lim}$ actually increases with measurement time. The rise of $SNR_{lim}$ is compensated by a decrease of noise mean value $N$. This ensures better detection sensitivity for longer measurements. For the two measurements times and gate interval $\tau = 1$ ms, the detectable limit levels are $SNR_{lim}(3\ s, 1\ ms) = 3.2\pm0.3$ and $SNR_{lim}(300\ s, 1\ ms) = 4.0\pm0.2$ with corresponding detectable laser frequency deviations $A_{lim} = 86\pm8$ kHz respectively $A_{lim} = 10.8\pm0.5$ kHz. In addition, for comparison with the previous paragraph and measurement times 5 and 500 s, the corresponding detectable frequency deviations are $A_{lim} = 68\pm5$ kHz s and $A_{lim} = 8.6\pm0.4$ kHz, respectively.

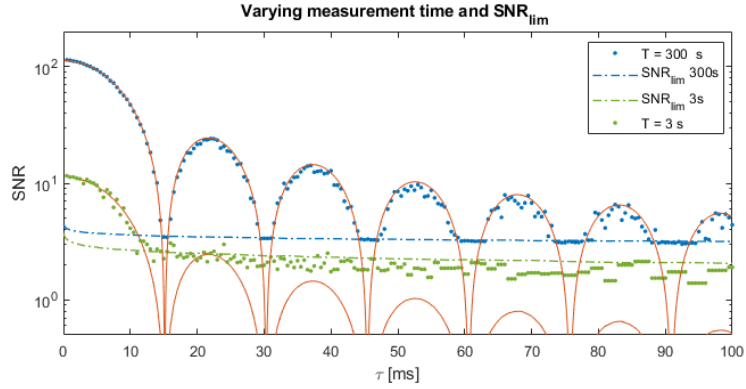

Fig. 6. Signal-to-noise ratio data for 3 s and 300 s measurement of fluorescence with harmonically modulated analyzed laser detuning $\Delta_{866}$. Modulation frequency and frequency deviation are fixed to 66 Hz and 300 kHz, respectively. The signal amplitudes are found as the maximum of the whole frequency spectrum except DC. The full circles show the measured data, red lines depict the corresponding simulated values. Dashed lines represent numerical limits given by simulated signal with Poisson distribution $SNR_{lim}$.

## 4.4 Measurement point

We have further searched for optimal measurement point by characterization of the reciprocal dependence of the $SNR$ on the count rate $R$, as shown in Fig. 7. Measurements of $SNR(R)$ for two modulation frequencies $f_m = 12$ Hz and $f_m = 120$ kHz with $A = 300$ kHz are realized along the dark resonance spectrum by changing the frequency detuning measurement point $\Delta_M$. Gate interval $\tau$ is chosen such that $sinc(f_m,\tau) > 0.99$. At each measurement point, the average $SNR$ of ten measurements with $T = 10$ is compared with theoretical $SNR$ calculated using the slope parameter *m* obtained directly from the measured fluorescence spectrum. Furthermore, the ratio between the measured results of the two $f_m$ sets shows a decrease of *SNR* for the signal with $f_m > R$ and that this decrease is proportional to $\sim 1/R$.

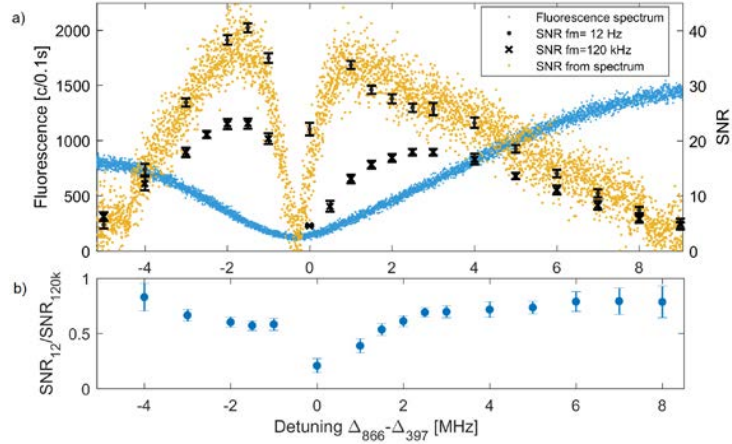

Fig. 7. a) The comparison of achievable signal-to-noise ratios for frequency analysis with various reference laser detuning. The detected fluorescence rates across dark resonance are shown as blue dots and the measured signal-to-noise data for 20 measurement points are shown for two modulation frequencies $f_m$ of 12 Hz and 120 kHz depicted as black dots and crosses, respectively. The yellow dots are theoretical *SNR* data calculated from the measured dark resonance. Plot b) shows the ratio between measured *SNR* for the two modulation frequencies.

## 5. Results and Discussion

We have proposed and implemented a method for the estimation of optical frequency spectrum using the time resolved measurement of light scattered from a single trapped ion with a sensitivity enhanced by the excitation on the slope of dark resonance. Fluorescence rate sensitivity to laser frequency deviation *A* is given by the slope function of the dark resonance *m*, which depends on the parameters of applied magnetic and laser fields. The measurement point $\Delta_{M1}$ stands on the slope, whose length in the axis of frequency detuning has been set by the parameters of excitation lasers approximately to 8 MHz. This gives us the fundamental upper limit of the presented method on the frequency deviation and modulation frequency. According to the Carson rule for frequency modulation bandwidth with 98% of modulation energy [22], combined frequency deviation and modulation frequency has to satisfy the condition $2(A+f_m) \approx 8$ MHz, this condition was well satisfied for all presented measurements. The smallest detectable modulation amplitudes are limited by the amount of fluorescence noise. In terms of signal to noise ratio, this limit is defined in two approaches; when the signal part is taken from real data at specific frequency – $SNR_{pl}$ and when the signal part is taken in the whole FFT spectrum of simulated data – $SNR_{lim}$. Specifically for measurements at measurement point $\Delta_{M1}$ the empirical limit is measured as $SNR_{pl} = 1.8 \pm 0.4$, which corresponds to the lowest detectable frequency modulation $A_{pl} = 38 \pm 8$ kHz for $T = 5$ s or $A_{pl} = 3.8 \pm 0.8$ kHz for $T = 500$ s. The simulated limit scales up with measurement time, however the mean level of amplitudes of noise frequency spectrum *N* is proportional to $\sqrt{R\tau/T}$ thus the lowest detectable frequency deviation scales down with longer measurement times. The corresponding detectable frequency deviations in our setup are $A_{lim} = 68 \pm 5$ kHz for $T = 5$ s and $A_{lim} = 8.6 \pm 0.4$ kHz for $T = 500$ s.

The detectable modulation frequency should be, according to the presented theory Eq. (5), unlimited. However, the measurement results show a decrease of $SNR(f_m, \tau)$ for high modulation frequencies. The decrease has been observed for modulation frequencies higher than the photon count rate $R$, where signals start to be undersampled. Comparisons of *SNR* for two modulation frequencies $f_m$ = 12 Hz and $f_m$ = 120 kHz measured at 20 points along the dark resonance indicate a correlation of the decrease with fluorescence intensity, although the theoretical simulation shows that the undersampling itself should not limit the observable modulation frequency detection bandwidth. Simulation results did not show any decrease of *SNR*, even for an order of magnitude higher modulation frequencies observed with the same photon rate. Other possible effects, including leaking of the spectral modulation sidebands out of the resonance slope or the frequency response of the AOM, were also investigated and do not explain the observable decrease of SNR. Thus, at this point, we leave the attainable bandwidth limit of the presented method, as well as the studies if possible excitation of the motional sidebands within the employed Raman excitation scheme might play a role, for further investigation.

The atomic level scheme that enables observation of dark-states in emitted fluorescence is occurring among vast majority of species commonly employed in ion trap experiments. Different branching ratios of the decay constants leads to different spectral shapes of the fluorescence. In general, the optimal spectral shape will depend on experimental parameters such as detuning of a cooling laser, laser intensities and strength and direction of applied magnetic field. A lower intensity and lower detuning of the cooling laser should always lead to narrower resonances and higher fluorescence gradients. On the other hand, such adjustment decreases the spectral width of dark resonances thus limiting spectral bandwidth of the method, also decreases fluorescence rate, and leads to unstable Doppler cooling performance. We provide a simplified comparison of $^{40}Ca^+$ with other frequently employed alkaline earth metal ions, such as $^{88}Sr^+$ and $^{138}Ba^+$ by estimating the values of fluorescence gradients on the same lambda scheme incorporating $S_{1/2}$, $P_{1/2}$ and $D_{3/2}$ levels. This is done by simulating dark resonance spectra, using optical Bloch equations and the same experimental parameters as described in section 2. For $^{40}Ca^+$ the highest gradient is on transition between Zeeman states $|S_{1/2},-1/2\rangle$ and $|D_{3/2},-1/2\rangle$ with $m_{(-1/2,-1/2)}$ = 2.3 counts·s$^{-1}$·kHz$^{-1}$. For $^{88}Sr^+$ it is $m_{(-1/2,-1/2)}$ = 2.4 counts·s$^{-1}$·kHz$^{-1}$ and for $^{138}Ba^+$ the highest gradient is on transition between states $|S_{1/2},+1/2\rangle$ and $|D_{3/2},+1/2\rangle$ with value $m_{(+1/2,+1/2)}$ = 2.6 counts·s$^{-1}$·kHz$^{-1}$. In practice, a better Doppler cooling performance and smaller broadening of resonances by motion for the heavier elements should increase the attainable gradients and higher detection efficiency at strontium's 422 nm and barium's 493 nm wavelengths can enhance the sensitivity of the method too.

The presented method of optical frequency analysis has been verified in a range of experimental parameters. The observed sensitivity to frequency deviations and achievable spectral bandwidths, which are in good agreement with theoretical model, are already sufficient for a large range of interesting applications in the optical spectral analysis [23]. Besides the techniques employing optical frequency combs, it offers an alternative method for frequency analysis of spectral noise of two frequencies of very distant lasers, which can excite transitions in atomic probes sharing a common level. The method can be extended to analysis of complementary experimental platforms for atomic trapping in optical lattices or tweezers. Another natural application of the presented scheme corresponds to analysis and phase locking of the two lasers for the purposes of coherent operations on Raman transitions incorporating two disparate wavelengths [7]. In addition, the method can be directly applied to spectral noise analysis of magnetic field seen by the ion. This simple sensing scheme with spatial resolution on the level of tens of nm can be beneficial when the knowledge of spatial dependence of the magnetic field within the Paul trap is of interest. Such knowledge is essential e.g. for quantum algorithms where multiple ions are stored at different locations of a segmented Paul trap and accumulate undesired position-dependent phases [24]. In comparison with traditional spectroscopy on narrow transitions or with direct analysis on dark resonances, this method can provide complete information of magnetic field change in time. The presented lowest detectable laser frequency deviations for $T$ = 500 s would allow for detecting magnetic field deviations of $B_{pl}$= 1.9±0.3 mG and $B_{lim}$= 2.8±0.1 mG in the corresponding frequency range. The sensitivity of the detectable fluorescence rate to the analyzed probe frequency deviations can be further improved by increasing the overall fluorescence detection efficiency, which has been in our case limited mostly by numerical aperture of the collection optics to 2% of the full solid angle. The other feasible option is to increase the number of trapped ions. Both approaches will linearly enhance fluorescence intensity and thus the steepness of the dark resonance slope.

**Funding**

Grant Agency of the Czech Republic (19-14988S); European Metrology Programme for Innovation and Research (EMPIR programme co-financed by the participating States and from European Union's Horizon 2020 research and innovation programme) (17FUN07 CC4C); Ministry of Education, Youth and Sports of the Czech Republic (CZ.02.1.01/0.0/0.0/16_026/0008460 and LO1212); Czech Academy of Sciences (RVO: 68081731); European Commission (ALISI No. 394 CZ.1.05/2.1.00/01.0017).

**Disclosures**

The authors declare no conflicts of interest.

**References**

1. C. Cohen-Tannoudji and D. Guéry-Odelin, *Advances in atomic physics : an overview* (World Scientific, Singapore ; Hackensack, NJ, 2011), pp. xxv, 767 p.
2. "Atomic physics: precise measurements and ultracold matter," Choice: Current Reviews for Academic Libraries **51**, 1846-1846 (2014).
3. A. Predojevic and M. W. Mitchell, "Engineering the Atom-Photon Interaction: Controlling Fundamental Processes with Photons, Atoms and Solids," Nanopt Nanophoto, 1-405 (2015).
4. A. D. Ludlow, M. M. Boyd, J. Ye, E. Peik, and P. O. Schmidt, "Optical atomic clocks," Rev Mod Phys **87**, 637-701 (2015).
5. D. Hayes, D. N. Matsukevich, P. Maunz, D. Hucul, Q. Quraishi, S. Olmschenk, W. Campbell, J. Mizrahi, C. Senko, and C. Monroe, "Entanglement of Atomic Qubits Using an Optical Frequency Comb," Phys Rev Lett **104**(2010).
6. A. G. Paschke, G. Zarantonello, H. Hahn, T. Lang, C. Manzoni, M. Marangoni, G. Cerullo, U. Morgner, and C. Ospelkaus, "Versatile Control of Be-9(+) Ions Using a Spectrally Tailored UV Frequency Comb," Phys Rev Lett **122**(2019).
7. C. Solaro, S. Meyer, K. Fisher, M. V. Depalatis, and M. Drewsen, "Direct Frequency-Comb-Driven Raman Transitions in the Terahertz Range," Phys Rev Lett **120**(2018).
8. M. Collombon, C. Chatou, G. Hagel, J. Pedregosa-Gutierrez, M. Houssin, M. Knoop, and C. Champenois, "Experimental Demonstration of Three-Photon Coherent Population Trapping in an Ion Cloud," Phys Rev Appl **12**(2019).
9. H. Schnatz, B. Lipphardt, J. Helmcke, F. Riehle, and G. Zinner, "First phase-coherent frequency measurement of visible radiation," Phys Rev Lett **76**, 18-21 (1996).
10. S. W. Kim, "Combs rule," Nat Photonics **3**, 313-314 (2009).
11. S. T. Cundiff, J. Ye, and J. L. Hall, "Optical frequency synthesis based on mode-locked lasers," Rev Sci Instrum **72**, 3749-3771 (2001).
12. R. Holzwarth, T. Udem, T. W. Hansch, J. C. Knight, W. J. Wadsworth, and P. S. J. Russell, "Optical frequency synthesizer for precision spectroscopy," Phys Rev Lett **85**, 2264-2267 (2000).
13. M. Collombon, G. Hagel, C. Chatou, D. Guyomarc'h, D. Ferrand, M. Houssin, C. Champenois, and M. Knoop, "Phase transfer between three visible lasers for coherent population trapping," Opt Lett **44**, 859-862 (2019).
14. E. Arimondo, "Coherent population trapping in laser spectroscopy," Prog Optics **35**, 257-354 (1996).
15. I. Siemers, M. Schubert, R. Blatt, W. Neuhauser, and P. E. Toschek, "The Trapped State of a Trapped Ion - Line Shifts and Shape," Europhys Lett **18**, 139-144 (1992).
16. J. Rossnagel, K. N. Tolazzi, F. Schmidt-Kaler, and K. Singer, "Fast thermometry for trapped ions using dark resonances," New J Phys **17**(2015).
17. C. Lisowski, M. Knoop, C. Champenois, G. Hagel, M. Vedel, and F. Vedel, "Dark resonances as a probe for the motional state of a single ion," Appl Phys B-Lasers O **81**, 5-12 (2005).
18. D. Reiss, K. Abich, W. Neuhauser, C. Wunderlich, and P. E. Toschek, "Raman cooling and heating of two trapped Ba+ ions," Phys Rev A **65**(2002).
19. J. Eschner, G. Morigi, F. Schmidt-Kaler, and R. Blatt, "Laser cooling of trapped ions," J Opt Soc Am B **20**, 1003-1015 (2003).

20. R. Loudon, *The quantum theory of light,* 3rd ed., Oxford science publications (Oxford University Press, Oxford ; New York, 2000), pp. ix, 438 p.
21. N. Scharnhorst, J. B. Wubbena, S. Hannig, K. Jakobsen, J. Kramer, I. D. Leroux, and P. O. Schmidt, "High-bandwidth transfer of phase stability through a fiber frequency comb," Opt Express **23**, 19771-19776 (2015).
22. J. R. Carson, "Notes on the theory of modulation," P Ire **10**, 57-64 (1922).
23. N. Picque and T. W. Hansch, "Frequency comb spectroscopy," Nat Photonics **13**, 146-157 (2019).
24. T. Ruster, H. Kaufmann, M. A. Luda, V. Kaushal, C. T. Schmiegelow, F. Schmidt-Kaler, and U. G. Poschinger, "Entanglement-Based dc Magnetometry with Separated Ions," Phys Rev X **7**(2017).